\documentclass[preprint,nofootinbib]{revtex4}
\usepackage{amsfonts}
\usepackage{amsmath}
\usepackage{graphicx}

\begin{document}

\title{Multiple scattering and PXR: kinematical suppression of
multiple scattering influence on PXR and dynamical shift of
diffraction peaks}
\author{Konstantin Batrakov}
\affiliation{Institute for Nuclear Problems, Belarus State
University.}

\begin{abstract}
One feature of passing  relativistic electron through a single
crystal is emission of parametric X-ray radiation (PXR) due to
diffraction on a crystallographic planes. As was shown in
\cite{1}, PXR and coherent bremsstrahlung (CBS) take place for
non-relativistic (or slow relativistic) electrons too. In this
case emission spectrum consists of number of peaks corresponding
to crystal reflexes in background of incoherent bremsstrahlung.
These peaks are distinguished only for thin crystals, because the
influence of electron multiple scattering on radiation increases
with the crystal thickness and, as a result, smoothes out peaks.
Simulation of PXR and CBS emission and considering of multiple
scattering influence on spectra are carried out in present work.
The optimal geometry with suppressed influence of multiple
scattering is considered. Optimal geometry corresponds to the
following condition
$$
\cos \psi =1/\sqrt{1+\left( \frac{\beta _{u}\sin \theta }{1-\beta
_{u}\cos \theta }\right) ^{2}}.
$$
Here $\psi$ is inclination angle of electron velocity vector
relative to the reciprocal lattice vector corresponding to the
reflex under consideration. $\beta_u=u/c$, where $u$ is average
velocity of electron in beam , $\theta$ is the angle between
velocity and wave vector of emitted photon. Optimal geometry gives
possibility to use more thick crystals for PXR generation. The
dynamical shift of PXR peaks which is one more feature of multiple
scattering influence is predicted and discussed.
\end{abstract}

\maketitle


\section{Introduction}

PXR is the radiation from a charged particle due to the photon
diffraction on the  planes of a single crystal. It was shown both
theoretically and experimentally  and citation therein) that PXR
from a relativistic electron ($E \geq 50~\mathrm{MeV}$) led to
very high spectral intensity within the narrow angular and
spectral range ($\Delta \theta \sim \Delta \omega/\omega \sim
10^{-3}$).

Work \cite{1} discussed and explained diffraction peaks produced
experimentally in \cite{2,3,4} for non-relativistic electrons ($E
\sim 50 \div 100 ~\mathrm{keV}$) on the basis of interference
between PXR and CBS.
    But the non-relativistic electron beam
passing through the crystal is under the strong action of Coulomb
multiple scattering. It action is so great that the electron beam
becomes completely  isotropic after passing through the crystal
with the thickness of order of several microns. So, the PXR and
CBS is lost in the background of incoherent bremsstrahlung. For
observation of PXR and CBS peaks the crystal must be sufficiently
thin ($\sim 0.1~\mu\textrm{m}$).

In the present paper influence of multiple scattering on PXR and
CBS peaks is studied. The features of PXR and CBS such as
dynamical peak frequency shifts and kinematical suppression of
multiple scattering are derived and discussed.

\section{Account of multiple scattering}
\label{sect:expressions}

 Corresponding to Feranchuck \cite{1}, the
spectral-angular distribution of emitted quanta has form:
\begin{eqnarray}
d^{2}N_{\mathbf{n}\omega }^{(s)}=\frac{e^{2}\omega }{4\pi \hbar c^{3}}\sum_{%
\mathbf{g}}\left\vert \mathbf{v}_{0}\mathbf{E}_{g_{s}}-\frac{e}{m}\frac{\mathbf{e}_{s}%
\mathbf{g}}{\mathbf{g}\mathbf{v}_{0}}U_{g}-\frac{e}{m}\mathbf{v}_{0}\frac{\mathbf{k}\mathbf{g%
}}{(\mathbf{g}\mathbf{v}_{0})^{2}}U_{g}\right\vert ^{2} \times \nonumber \\
\left\vert \frac{\exp [i((%
\mathbf{k}+\mathbf{g})\mathbf{v}_{0}-\omega )t]-1}{(\mathbf{k}+\mathbf{g})\mathbf{v}%
_{0}-\omega }\right\vert ^{2}d\omega d\Omega.  \label{eq1}
\end{eqnarray}
Here $\mathbf{E}_{g_{s}}=\chi _{g}\left(c^{2}\mathbf{k}%
_{g}(\mathbf{g}\mathbf{e}_{s})-\omega
^{2}\mathbf{e}_{s}\right)/\left(k_{g}^{2}c^{2}-\omega
^{2}\right)$,$U_g=$, $\chi_g$ is crystal polarizabilities,
$\mathbf{e}_s$ is photon polarization vector, $\mathbf{g}$ is
reciprocal lattice vector, $\mathbf{v}_0$ is electron velocity,
$e,m$ are electron electrical charge and mass. $t$ is time of
electron passing through the crystal, $\mathbf{k}$ and $\omega$
are
wave vector and frequency of emitted photons. \\
Electron beam under action of multiple scattering acquires
velocity spread, therefore (\ref{eq1}) should be averaged over
this spread:
\begin{eqnarray}
d^{2}N_{\mathbf{n}\omega }^{(s)} &=&\frac{e^{2}\omega }{4\pi \hbar
c^{3}}\int \frac{d\varphi_0 }{2\pi }d\theta _{0}\theta
_{0}f(\theta _{0})\sum_{\mathbf{g}}\left\vert
\mathbf{v}_{0}\mathbf{E}_{g_{s}}-\frac{e}{m}\frac{\mathbf{e}_{s}\mathbf{g}}{\mathbf{g}%
\mathbf{v}_{0}}U_{g}-\frac{e}{m}\mathbf{v}_{0}\frac{\mathbf{k}\mathbf{g}}{(\mathbf{g}\mathbf{%
v}_{0})^{2}}U_{g}\right\vert ^{2}\times  \nonumber \\
&&\left\vert \frac{\exp [i((\mathbf{k}+\mathbf{g})\mathbf{v}_{0}-\omega )t]-1}{(%
\mathbf{k}+\mathbf{g})\mathbf{v}_{0}-\omega }\right\vert
^{2}d\omega d\Omega. \label{spread}
\end{eqnarray}
In (\ref{spread}) $f(\theta _{0})$ is distribution function of
electron beam ($\theta _{0}$ is angle relative to vector of
average velocity). Distribution function will be used in gaussian
form
\begin{equation}
f(\theta _{0})=\frac{2\exp \left\{ -\frac{ \displaystyle\theta
_{0}^{2}}{\displaystyle \theta _{s}^{2}}\right\} }{\theta _{s}^2}.
\label{distribution}
\end{equation}
In (\ref{distribution}) $\theta _{s}$ is scattering angle the
square of which  equals to $\theta _{s}^{2}=16\pi
n_{0}Z(Z+1)r_{e}^{2}\frac{\displaystyle(E+1)^{2}}{\displaystyle
E^{2}(E+2)^{2}}L_{K}L_e$. $L_e$ is an electron path in crystal
\cite{5}, $Z$ is the crystal atom number, $E=T/(m c^2)$, $T$ is
electron kinetic energy, $L_k$ is Coulomb logarithm, $n_0$ is atom
density. Typical value of scattering angle for $Si$ at path length
$1~\mu\mathrm{m}$ is $\theta _{s}\sim 0.4~\mathrm{rad}$. In
(\ref{spread})  angular dispersion is taken into account only.
Energy dispersion and energy losses exerts significantly less
influence on radiation at considered energies (electron energy
losses on ionization in $Si$ is of order of $100$ eV at electron
energy $E=100$ keV and crystal thickness $1~\mu$m).

Line width of electron spontaneous emission without accounting the
multiple scattering can be estimated as $\Delta \omega/\omega \sim
1/(k L_e)$ (if  $L_e$ is less than absorption length) and it gives
$\Delta \omega/\omega \sim 10^{-2}$ for crystal thickness $\sim
0.1~\mu \mathrm{m}$. Therefore, scattering angle for crystal $\ge
0.1~\mu \mathrm{m}$ is greater than spontaneous emission line
width. In this case we can use in (\ref{spread}) the following
approximation
\begin{equation}
\left\vert \frac{\exp [i((\mathbf{k}+\mathbf{g})\mathbf{v}_{0}-\omega )t]-1}{(\mathbf{k}%
+\mathbf{g})\mathbf{v}_{0}-\omega }\right\vert ^{2}=2\pi \delta \left( \omega -(%
\mathbf{k}+\mathbf{g})\mathbf{v}_{0}\right) \frac{1-\exp \left\{
-k_{z}^{"}L\right\} }{k_{z}^{"}v_{0z}}, \label{approx}
\end{equation}
where $k_{z}^{"}$ is imaginary part of wave vector component
$k_z$, $L$ is crystal thickness. Integrating (\ref{spread}) over
$\theta_0$ gives
\begin{eqnarray}
d^{2}N_{\vec{n}\omega }^{(s)}=\frac{e^{2}1-\exp \left\{
-k_{z}^{"}L\right\}}{4\pi \hbar c^{3}k_{z}^{"}u_{z}\pi \theta
_{s}^{2}}\sum_{\vec{g}}\left\vert A\right\vert ^{2}\int ^{'}
d\varphi _{0}\frac{1 }{\sqrt{\left[ \ \tan (\theta _{g })\cos
(\varphi -\varphi _{0})\right] ^{2}-2\Delta_g }}\times \nonumber
\\
\times \left\{ \left( \tan (\theta _{g })\cos (\varphi -\varphi _{0})+%
\sqrt{\left[ \ \tan (\theta _{g })\cos (\varphi -\varphi
_{0})\right] ^{2}-2\Delta_g }\right) \right. \nonumber\\
 \exp \left[ -\left(
\tan (\theta _{g })\cos (\varphi -\varphi _{0})+\sqrt{\left[ \
\tan (\theta _{g })\cos (\varphi -\varphi _{0})\right]
^{2}-2\Delta_g }\right) ^{2}/\theta _{s}^{2}\right]
+ \label{midspectral} \\
 +\left( \tan (\theta _{g })\cos (\varphi -\varphi _{0})-\sqrt{%
\left[ \ \tan (\theta _{g })\cos (\varphi -\varphi _{0})\right]
^{2}-2\Delta_g }\right) \nonumber \\
\left.
 \exp \left\{ -\left( \tan (\theta _{g
})\cos (\varphi -\varphi _{0})-\sqrt{\left[ \ \tan (\theta _{g
})\cos (\varphi -\varphi _{0})\right] ^{2}-2\Delta_g }\right)
^{2}/\theta _{s}^{2}\right] \right\} d \omega \Omega. \nonumber
\end{eqnarray}
Here $\varphi,\varphi_0$ are the polar angles corresponding to
wave vector and electron velocity vector. $\theta_{g}$ is angle
between vector $\mathbf{k + g}$ and average velocity vector
$\mathbf{v}_0$.
$\Delta_g=\left(\omega-\mathbf{(k+g)v_0}\right)/\omega$ is
parameter of detuning from synchronism. The prime over the
integral means that integation of terms containing $z_{1}=\tan
(\theta _{g })\cos (\varphi -\varphi _{0})+\sqrt{\left[ \ \tan
(\theta _{g })\cos (\varphi -\varphi _{0})\right] ^{2}-2\Delta
_{g}}$ and $z_{2}=\tan (\theta _{g })\cos (\varphi -\varphi
_{0})-\sqrt{\left[ \ \tan (\theta _{g })\cos (\varphi -\varphi
_{0})\right] ^{2}-2\Delta }$ is fulfilled in regions of
$\varphi_0$, where $z_1$ and $z_2$ are positive. For convenience
let's perform substitution $x=\cos (\varphi - \varphi_0)$. That
gives the integral in the form
\begin{equation}
\begin{array}{c}
2\int^{\prime }\frac{\displaystyle dx}{\displaystyle\pi \theta
_{s}^{2}\sqrt{\left[ \ \tan (\theta
_{g })x\right] ^{2}-2\Delta _{g}}\sqrt{1-x^{2}}} \\
\left\{ \left( \tan (\theta _{g })x+\sqrt{\left[ \ \tan (\theta _{g})x%
\right] ^{2}-2\Delta _{g}}\right) \exp \left[ -\left( \tan (\theta
_{g})x+\sqrt{\ \tan ^{2}(\theta _{g })x^{2}-2\Delta _{g}}\right)
^{2}/\theta
_{s}^{2}\right] +\right.  \\
\left. \left( \tan (\theta _{g })x-\sqrt{\left[ \ \tan (\theta _{g })x%
\right] ^{2}-2\Delta _{g}}\right) \exp \left[ -\left( \tan (\theta
_{g })x-\sqrt{\ \tan ^{2}(\theta _{g })x^{2}-2\Delta _{g}}\right)
^{2}/\theta _{s}^{2}\right] \right\}.
\end{array}
\label{integral}
\end{equation}
In (\ref{integral}) integration is fulfilled over the regions of
$-1...1$, in which $z_1$ and $z_2$ are positive. As was mentioned
above even for crystal thickness $L\sim0.1~\mu \mathrm{m}$
scattering angle exceeds spontaneous emission line width
$\theta_s>1/(k L_e)$. As a result, emission line is widened and
spectral-angular brightness reduces. It is very desirable to
decrease the influence of multiple scattering. Such possibility
really exists. So $\mathbf{k_g v_0}=k_g v_0 \cos (\theta_g)$, then
due to electron beam spreading $\cos (\theta_g+\theta_s)\approx
\cos (\theta_g)-\sin (\theta_g)\theta_s-\cos (\theta_g) \theta_s^2
/2$. Therefore if $\sin (\theta_g)=0$, the terms corresponding to
detuning from synchronism ($\omega - \mathbf{k_g v_0}\approx 0 $)
are proportional to $\theta_s^2$, not to $\theta_s$. So, when
$\theta_s\ll 1$, the geometry with $\sin \theta_g=0$ is
preferable. As will be shown bellow in section \ref{sect:simul}
this geometry actually can essentially increase (on several
orders) spectral-angular brightness of emission.

Corresponding to (\ref{approx}) the emission peaks of single
electron has frequencies
\begin{equation}
\omega_g =\frac{\mathbf{g v}_{0}}{1-\mathbf{n v_0}/c}.
\label{peaks}
\end{equation}
Here $\mathbf{n}$ is a unit vector having direction of wave
vector. So, position of these peaks depends on reciprocal vector
$\mathbf{g}$ and on angle of photon observation. Therefore,
spectral distribution measured by detector at fixed angle is
looked as the set of peaks on the background of incoherent
bremsstrahlung. The peaks position changes as with the observation
angle change so with crystal rotation. Therefore, emission
spectrum is easy tuned.
  Due to multiple scattering electron beam acquires
velocity spread, which leads to frequency spread in correspondence
to expression (\ref{peaks}). First term of (\ref{eq1})
corresponding to PXR depends on polarizability $\chi_g$.
$\vert\chi_g \vert^2\sim \omega^{-4}$, therefore this term
increases with frequency decrease. On the other hand, CBS grows
with frequency increase. Therefore, if PXR exceeds CBS, then
diffraction peak drift in soft spectral range. On the contrary, if
CBS essentially exceeds PXR then diffraction peak shifts in region
with larger frequency.
 This dynamical frequency
shift will be demonstrated in \ref{sect:simul}.

\section{Discussion and estimations.}

\label{sect:simul}
\begin{figure}[htb]
\centering {\hspace{-1.2 cm}
\begin{minipage}[t]{7cm}
{\includegraphics[ scale=1]{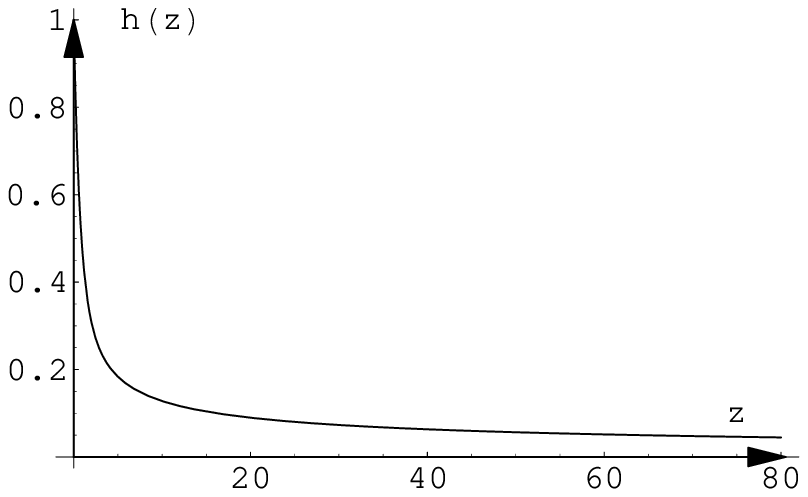}}

\centerline{a)}
\end{minipage}
\qquad  \qquad
\begin{minipage}[t]{7cm}
{\includegraphics[ scale=1]{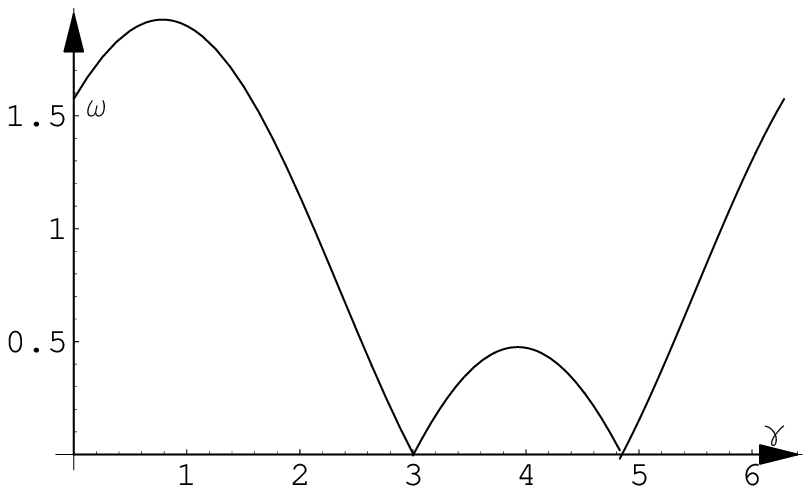}}

\centerline{b)}
\end{minipage}
} \caption{ a) Graphic of function $\mathrm{h}(z)$ dependence on
it argument; b) Frequency dependence on $\gamma$. Si [0 0 1],
reflex (1 1 1). Electron energy $E=75$ KeV.}\label{fig:gammarot}
\end{figure}

 Let's study some common features of spectral
angular distribution accounting the multiple scattering. If
$\tan\theta_g\geq 1$, then following estimation of
(\ref{integral}) can be made
\begin{equation}
\frac{2}{\theta _{s}^{2}}\exp \left\{ -\frac{2\tan ^{2}(\theta _{g })}{%
\theta _{s}^{2}}\right\} \mathrm{I}_0\left( \frac{2\tan
^{2}(\theta _{g })}{\theta _{s}^{2}}\right) +\frac{1}{\sqrt{\pi
}\tan (\theta _{g })\theta _{s}}\exp \left\{ -\frac{\Delta_g
^{2}}{\tan ^{2}(\theta _{g })\theta _{s}^{2}}\right\}.
\label{bigangle}
\end{equation}
Expression (\ref{bigangle}) is true if $\tan (\theta _{g
})\gg\Delta_g$ and $\tan (\theta _{g })\gg\theta_s$. The first
term of
(\ref{bigangle}) is proportional $h=\exp \left\{ -\frac{2\tan ^{2}(\theta _{g })}{%
\theta _{s}^{2}}\right\}$ $ \mathrm{I}_0\left(\frac{2\tan
^{2}(\theta _{g })}{\theta _{s}^{2}}\right)$ (dependence
$\mathrm{h}(z)$ on it argument is shown on
Fig.~\ref{fig:gammarot}a). It follows that spectral width of
diffraction peak $\delta\omega/\omega$ is proportional to
$\vert\tan (\theta _{g })\vert\theta _{s}$ and maximal brightness
grows with reducing this width. So, spectral brightness increases
with decreasing $\vert\tan (\theta _{g })\vert$ as was predicted
in \ref{sect:expressions} and kinematical suppressing of multiple
scattering influence takes place.

In opposite case $\vert\tan (\theta _{g })\vert=0$
(\ref{integral}) can be estimated as
\begin{equation}
\frac{2\exp \left\{ -\frac{\displaystyle 2\vert\Delta _{g}\vert}{\displaystyle\theta _{s}^{2}}\right\} }{%
\theta _{s}^{2}},  \label{nullangle}
\end{equation}
and spectral width $\delta\omega/\omega$ is proportional to
scattering angle squared. Expression (\ref{nullangle}) is true for
$\Delta_g\gg\tan\theta_g$.   (\ref{nullangle}) is applied for
large angle $\theta_g$  also. It follows  from (\ref{bigangle})
and (\ref{nullangle}) that when $\mathbf{k+g}\parallel
\mathbf{v}$, value of spectral intensity exceeds the spectral
intensity for $\tan \theta_g\gg \theta_s$ in $\sim \tan
\theta_g/\theta_s$ times. When using crystal with thickness for
which $\theta_s\gg 1$ it is generated bright and narrow spectral
line.

\begin{figure}[htb]
\centering {\hspace{-1.2 cm}
\begin{minipage}[t]{7cm}
{\includegraphics[ scale=0.8]{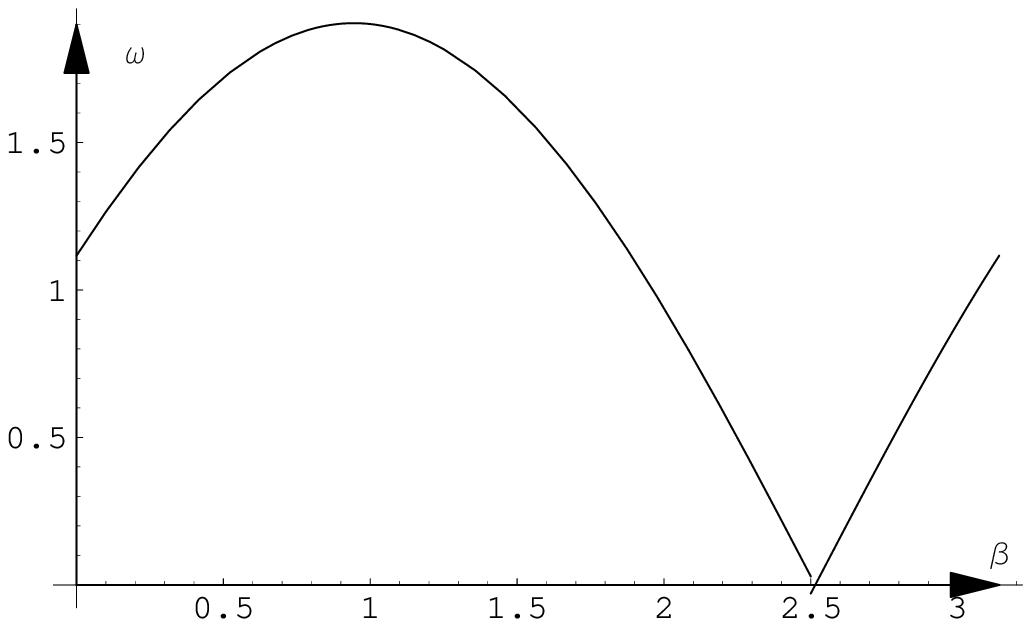}}

\centerline{a)}
\end{minipage}
\qquad  \qquad
\begin{minipage}[t]{7cm}
{\includegraphics[ scale=1]{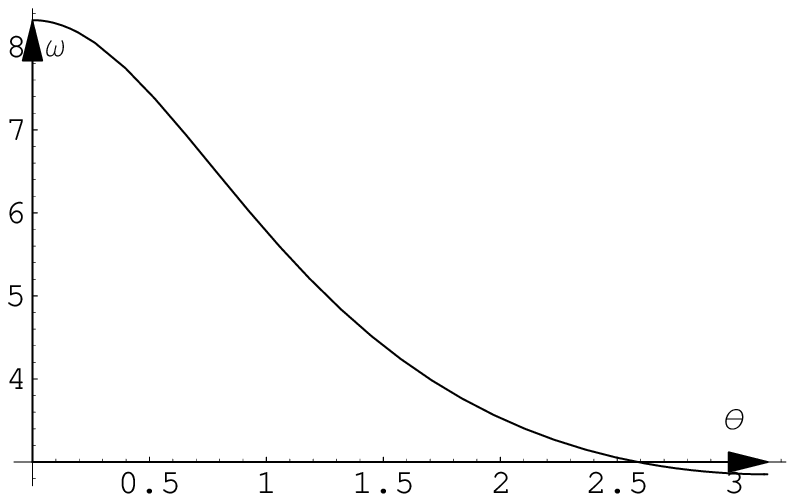}}

\centerline{b)}
\end{minipage}
} \caption{ a) Frequency dependence on $\beta$, Si [0 0 1], reflex
(1 1 1). Electron energy $E=75$ KeV.; b)Frequency dependence on
observation angle $\theta$, Si [0 0 1], reflex (0 0 4). Electron
energy $E=75$ KeV.}\label{fig:observrot}
\end{figure}

Let us study frequency dependence of diffraction peaks on crystal
rotation angles. Rotation will be described by standard Euler
angles $\alpha , \beta, \gamma$ between normal to the crystal
surface and velocity vector (which is parallel to $Z$ axis). Here
$\alpha$ changing in the range $0...2\pi$ is the rotation angle
relative to vector $v$, $\beta$ changing in range $0...\pi$ is
rotation angle relative to new axis $(1~0~0)$ and $\gamma$
changing in the range $0...2\pi$ is the rotation angle relative to
the new $Z'$ axis. It is evident that peak frequency can't depend
on $\alpha$ because independence of scalar product $\mathbf{g v}$
on $\alpha$. Dependence of frequency on $\gamma$ and $\beta$ is
shown on Figs. \ref{fig:gammarot}b and \ref{fig:observrot}a.

It can be seen from figures that frequency changes in wide
spectral range due to crystal rotation. Frequency changes with
change of observation angle also. This is shown on Fig.
\ref{fig:observrot}b.
These figures of frequency dependence were built with the help of
the formula
\begin{equation}
\omega =\frac{2\pi \beta _{u}}{a}\frac{n_{3}\cos \beta +n_{2}\sin
\beta \cos \gamma +n_{1}\sin \beta \sin \gamma }{1-\beta _{u}\cos
\theta }, \label{obsrot}
\end{equation}
which follows from (\ref{peaks}) if crystal surface coincides with
crystallographic plane [$0~0~1$], $\beta_u=v_0/c$. $n_1,~n_2,~n_3$
are indices corresponding to concerned reflexes.

\begin{figure}
   \begin{center}
   \begin{tabular}{c}
   \includegraphics[height=7cm]{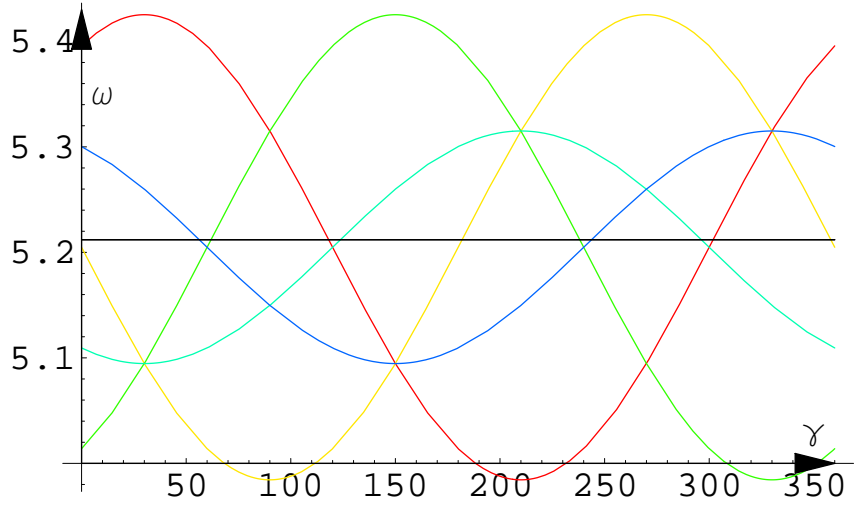}
   \end{tabular}
   \end{center}
   \caption[example]
   { \label{fig:splitting}
Splitting of reflex $k+l+m=7$ ($Si$ [1 1 1]) at nonzero $\beta=3$
$E=100$ KeV. Black line demonstrates dependence of peaks with
$k+l+m=7$ on $\gamma$ for zero inclination angle. Color curves
show frequency dependence of reflexes (5 1 1),(1 1 5), (1 5 1), (1
3 3), (3 1 3), (3 3 1) on $\gamma$ for inclination angle
$\beta=3$ degrees. }
   \end{figure}

Very effective  demonstration of orientation dependence is
splitting of reflexes frequencies. For example, all reflexes of
crystal [1 1 1] with same sum $k+l+m$ are degenerated at zero
angle of crystal inclination. This degeneration is disappeared if
we incline crystal on angle $\beta$. Dependence of frequency on
angle $\gamma$ is shown on Figure \ref{fig:splitting}. It is seen,
that for the case when crystal is not inclined, frequencies for
all reflexes with $k+l+m=7$ are the same and not change. But when
crystal is inclined on $\beta=3$ degrees, reflexes split and their
frequencies depend on $\gamma$. Let's note, that at different
values of angle $\gamma$ we shall see different number of reflexes
frequencies.

As was shown in (\ref{bigangle}) and (\ref{nullangle}), brightness
has maximal magnitude at $\theta_g\approx 0$. Let's study this
geometry in more detail. It follows from (\ref{peaks}) and
requirement of $\mathbf{k+g \parallel u}$ for $\theta_g = 0$ the
condition
\begin{equation}
\frac{\mathbf{g\beta }_{u}}{1-\mathbf{\beta_u
n}}\mathbf{n}+\mathbf{g\parallel \beta }_{u}  \label{condition}
\end{equation}
must be fulfilled. It follows from (\ref{condition}), that
component of vector $\mathbf{g}$ which is parallel to
[$\mathbf{k\times u}$]  must be zero. Besides, component of vector
in left hand side of (\ref{condition})  which is perpendicular to
$\mathbf{u}$ and lies in plane of $\mathbf{k}$ and $\mathbf{u}$
must be zero also. This leads to  expression for
$\psi=\mathbf{\widehat{g v}}$
\begin{equation}
\cos \psi =1/\sqrt{1+\left( \frac{\beta _{u}\sin \theta }{1-\beta
_{u}\cos \theta }\right) ^{2}}. \label{optimal}
\end{equation}
As it follows from (\ref{optimal}), the parameters of optimal
geometry depend on reflex, electron energy and observation angle.
Specific observation angle corresponds to specific crystal
rotation which gives maximal brightness for given reflex.

\begin{figure}
   \begin{center}
   \begin{tabular}{c}
   \includegraphics[height=7cm]{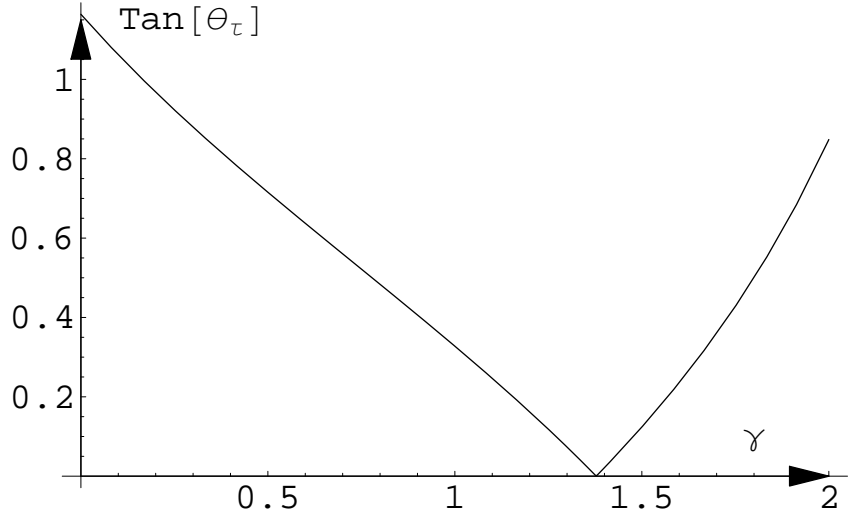}
   \end{tabular}
   \end{center}
   \caption[example]
   { \label{fig:tang}
Dependence of $\tan \theta_g$ on Euler angle $\gamma$, Si [0 0 1],
reflex (1 1 1). Electron energy $E=75$ KeV.}
   \end{figure}

Dependence of $\tan \theta_g$ on  Euler angle $\gamma$ is
demonstrated on Figure \ref{fig:tang}. This graphic is built at
fixed angles $\alpha=0, \beta=1.1 \pi/4, \theta=\pi/2-0.1$. Angle
between vector $\mathbf{k+g}$ and electron velocity vector
$\mathbf{v}$ $\theta_g$ is in vicinity to zero when $\gamma\approx
1.46$ for these parameters. It is evident that change of one angle
parameter changes other angular parameters  of target angular
position and observation angle corresponding to $\theta_g=0$.

\begin{figure}
   \begin{center}
   \begin{tabular}{c}
   \includegraphics[height=7cm]{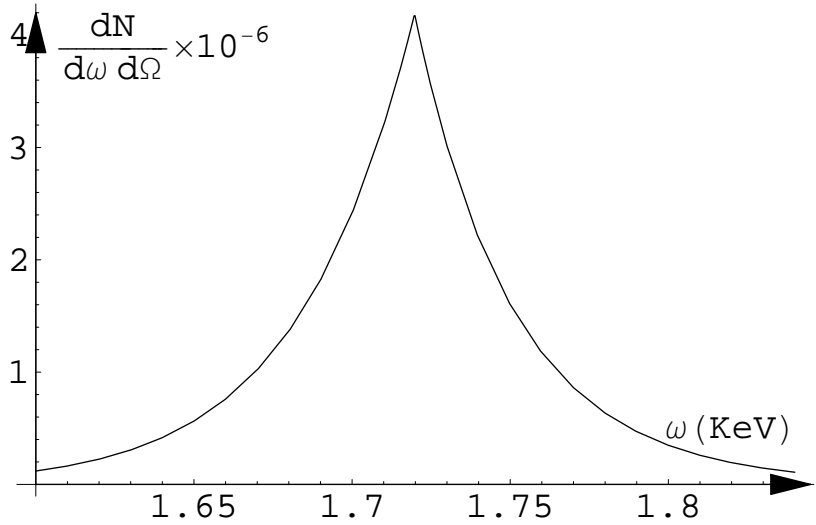}
   \end{tabular}
   \end{center}
   \caption[example]
   { \label{fig:optimanumber}
Optimal geometry. Frequency dependence of emitted photon spectral
angular density at $\theta_g\approx 0$. $\gamma=1.38$, $\theta =
\pi/2-0.1,~\alpha=0,~\beta=1.1\pi/4$. Si [0 0 1], reflex (1 1 1),
electron energy $E=75$ KeV.}
\end{figure}
\begin{figure}
   \begin{center}
   \begin{tabular}{c}
   \includegraphics[height=7cm]{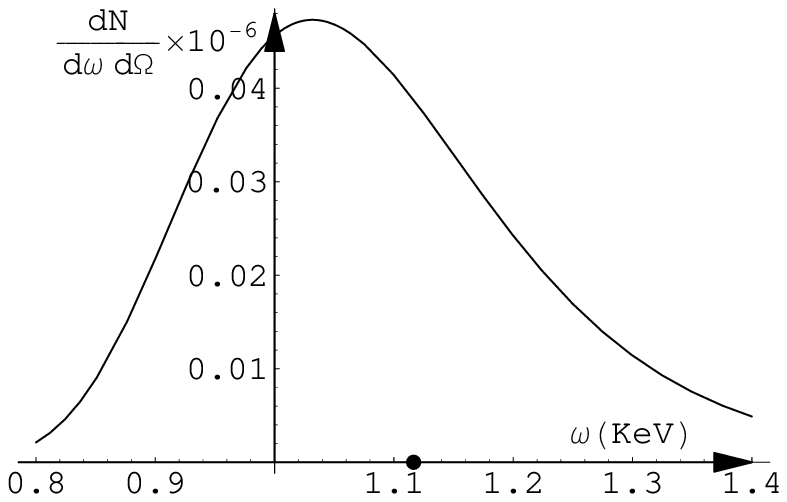}
   \end{tabular}
   \end{center}
   \caption[example]
   { \label{fig:nonopt}
Non-optimal geometry. Frequency dependence of emitted photon
spectral angular density . $\gamma=0$, $\theta =
\pi/2-0.1,~\alpha=0,~\beta=0$. Si [0 0 1], reflex (1 1 1),
electron energy $E=75$ KeV.Point shows central frequency of
diffraction peak without accounting of multiple scattering.
Dynamical frequency shift corresponds to difference of  this point
frequency
 and real peak central frequency. }
\end{figure}
Figure \ref{fig:optimanumber} corresponds to electron current
$1~\mu \mathrm{A}$ and crystal thickness $L\approx 0.1~\mu
\mathbf{m}$. Other parameters is on the Figure It can be seen,
that peak corresponding to optimal geometry  is very bright  and
it width is about of $0.05$ KeV. For comparison the dependence of
spectral-angular brightness on frequency for non-optimal geometry
is shown on Fig. \ref{fig:nonopt}. When the beam takes dispersion,
then electrons  which emit smaller frequencies will give  greater
contribution to emission due to increasing of polarizability at
frequency reducing. So, peaks will shift in more soft region.  On
Fig. \ref{fig:nonopt} point shows central frequency of diffraction
peak without accounting of multiple scattering. Dynamical
frequency shift corresponds to difference between these frequency.
Such behavior of peak central frequency  should be observed
experimentally.

The brightness corresponding to optimal geometry (Fig.
\ref{fig:optimanumber})  exceeds in $\sim 100$ times the
brightness for non-optimal one (Fig. \ref{fig:nonopt}).

One more method to suppress multiple scattering influence on
radiation is using of inclined geometry geometry for PXR emitted
photons observation. In the case when $\psi_1/(k \chi^{\prime
\prime}_0)\ll L$ ($\psi_1$ is the angle between wave-vector
$\mathbf{k}$ and crystal surface), the emission will be detected
from the electron which  path is of order of $\psi_1/(k
\chi^{\prime \prime}_0)$, not $\sim L$ or $\sim 1/(k\chi_0^{\prime
\prime})$. If $\psi_1/(k \chi^{\prime \prime}_0)$ is sufficiently
small, emission from such path is not disturbed by multiple
scattering. Combining kinematical suppression with inclined
observation geometry it is possible to produce very bright and
narrow peak of PXR radiation.

\section{CONCLUSION}
\label{conclusion}

Regulation of multiple scattering action on PXR (CBS) emission by
geometry choice gives possibility using relatively thick crystals
for observation of diffraction peaks. Specific reflex and specific
crystal orientation corresponds to specific observation angle and
frequency where kinematical suppression is appeard. Therefore,
experimental setup should have possibility of crystal target and
detector orientation in wide spectral range.

\subsection{Acknowledgments}
The work was fulfilled due to support ISTC: grant B-626.


\begin{thebibliography}{99}

\bibitem{1} I. D.~ Feranchuk, A.~ Ulyanenkov, J.~ Harada and J. C. H. Spence,
   Phys.\ Rev. A \ \textbf{62}, 4225 (2000). 


\bibitem{2} Y.S.~ Korobochko, V.F.~ Kosmach and V.I.~ Mineev,
 Sov. \ Phys. \ JETP,\  \textbf{21}, 834 (1965).




\bibitem{3} J.C.H.~ Spence and G.~ Reese,
    Acta \ Crystallogr. \ A, \textbf{42}, 577 (1986).



\bibitem{4} G.M.~ Reese, J.C.H.~ Spence and N.~ Yamamoto,
    Philos.\ Mag.\ A", \textbf{49}, 697 (1984).

\bibitem{5} V.S.~ Remizovich, D. B.~ Rogozkin and M. I.~ Ryazanov,
   \textquotedblleft Fluctuations of Free Path of Charged Particles",
    EnergoAtomIzdat, Moscow, (1988).


\end{thebibliography}
\end{document}